# Absolute Return Volatility

**JOHN COTTER***
University College Dublin

**Address for Correspondence:**
Dr. John Cotter,
Director of the Centre for Financial Markets,
Department of Banking and Finance,
University College Dublin,
Blackrock,
Co. Dublin,
Ireland.

E-mail. john.cotter@ucd.ie
Ph. +353 1 716 8900
Fax. +353 1 283 5482

June 9, 2005.

* The author would like to acknowledge the financial support from CPA and a UCD faculty grant.

# Absolute Return Volatility

**The use of absolute return volatility has many modelling benefits says John Cotter. An illustration is given for the market risk measure, minimum capital requirements.**

Volatility modelling is a key issue for the finance industry from an academic and practitioner perspective. This is understandable given the importance that volatility plays in risk management and the development of accurate risk measures. To illustrate, successful market risk management requires the use of accurate risk measures such as minimum capital requirements. These risk measures are underpinned the input of volatility estimates.[1]

An important key question thus arises: How do we obtain accurate volatility measures that can be used in market risk management? This paper addresses this by exploring the asymptotic and finite sample properties of absolute return volatility. Absolute return volatility is obtained by aggregating high frequency absolute returns into relatively low frequency, for example, daily volatility estimates. The use of these measures is illustrated by obtaining the commonly used market risk measure, minimum capital requirements. We advocate the use of absolute return volatility gives desirable time series properties and provides accurate measures of volatility.

We begin by noting that standard risk management practices postulate that market returns belong to a gaussian distribution. As we know this is not so and leads to inadequate risk measurement. For example, commonly cited deviations from normality in financial time series are the existence of fat-tails and of serial correlation in the volatility series, leading to a bias in minimum capital requirement estimates.[2] If however, volatility can be adequately modelled, the risk manager can filter out these properties from the returns series leading to a gaussian series. These standardised gaussian returns allow the risk manager to provide conservative and accurate risk measures.

---

[1] Minimum capital requirements represent reserves that are used to protect financial firms against losses arising from the volatility of their holdings (see Cotter, 2004; for a discussion).

[2] A key focus of many studies in market risk measurement is to explore alternative volatility processes. For instance, a number of conditional based approaches using Generalised Autoregressive Conditional Heteroskedastic (GARCH) and related univariate or multivariate process have been advocated (Brooks et al 2002). In addition, unconditional approaches that rely on separate risk measures for the upside and downside of a distribution have been supported such as the use of Extreme Value Theory (Longin 1996, 2000).

Absolute returns, overlooked in comparison to the use of squared returns, have many advantages in modelling volatility. First, absolute returns are robust in the presence of extreme or tail movements (Davidian and Carroll, 1987). Tail returns with their noted fat-tailed characteristic in financial time series are of particular importance in market risk management and in associated risk measures such as Value at Risk and minimum capital requirements. Second, accurate measures of unobservable latent volatility are obtained from absolute return volatility asymptotically through the theoretical framework of realised power variation. Moreover, absolute return volatility gives desirable finite sample properties that are applicable in practice for the risk manager. In particular, the properties match those found in market returns including serial correlation and by standardising the return series we eliminate these features.[3] Also, absolute return volatility measurement uses data with the highest frequency and this is beneficial in getting more precise estimates of risk measures (Merton, 1980).

The theoretical framework of realised power variation that underpins absolute return volatility is now outlined. This is followed by an illustration of the use of absolute return volatility in the calculation of minimum capital requirements for long and short trading positions on the FTSE100 futures contract.

*Realised power variation:*
One recent major innovation in the volatility literature has been the employment of realised power variation where realised volatility converges in probability to integrated volatility. Accurate model free volatility estimates are thus obtained. This theory relied on in the continuous time literature results in gaussian return innovations being a standard assumption of the pricing models presented.

The theoretical developments have evolved in conjunction with vast improvements in high frequency data allowing the continuous time framework to be realistically examined in a discrete context. The price process is assumed to follow Brownian motion and allows for accurate estimates of unobservable volatility at the limit. Discrete approximations of the price process using high frequency data have $r_{m,t} = p_t - p_{t-1/m}$ as the continuously compounded returns with m evenly spaced observations per day. Brownian motion is generalized to allow the volatility to be random but serially dependent exhibiting the stylized finding for financial return series of volatility clustering with fat-tailed unconditional distributions.[4]

---

[3] Also absolute return modelling has a number of other attractive features. For instance, absolute return modelling is more reliable than squared returns for the non-existence of a fourth moment commonly associated with financial returns (Mikosch and Starcia, 2000).

Volatility of this price process as measured by integrated volatility is unobservable. However, realised power variation that incorporates realised absolute variation, namely the sum of absolute realisations, $\sum |r_m|$, of a process captured at very fine intervals equate with integrated volatility. This theory of realised power variation given in Barndorff-Nielsen and Shephard (2003) and Barndorff-Nielsen et al (2003) extends the framework of quadratic variation presented for different square powers.[5] Thus for returns that are white noise and $\sigma^2_t$ with continuous sample paths, the limiting difference between the unobserved volatility estimate and the realised observed absolute variation is zero.

Barndorff-Nielsen and Shephard (2003) and Barndorff-Nielsen et al (2003) show that when the framework is for limiting intervals with $m \to \infty$, and with power variations, $0.5 > n < 3$, realised power variation converges in probability to integrated volatility.

$$p\lim_{m \to \infty} \left( \int_{t-H}^{t} \sigma^2_\tau d\tau - \sum_{j=1,...,m} |r_{m,t+j/m}| \right) = 0 \qquad (1)$$

Implying for m sampling frequency, the realized absolute variation is consistent with integrated volatility. Asymptotically the returns process scaled by realised power variation is normally distributed, N (0, 1).

---

[4] A number of semi-martingales can be utilised, and volatility modelling in this way allow for any number of characteristics documented for financial time series such as long memory and non-stationarity.

[5] The use of squared returns relying on quadratic variation has become a tour de force in the recent volatility literature with many studies completed. A flavour of the use of these related measures and a synopsis of the prevailing literature is in Andersen et al (2003). Similar to realised power variation the theory of quadratic variation implies that after assuming sample returns are white noise and $\sigma^2_t$ has continuous sample paths, the limiting difference between the unobserved volatility estimate and the observed realizations of the squared returns process is zero (Karatzas and Shreve (1991)).

Notwithstanding the derivation of the limiting distribution, our interest in the modelling process for risk measurement is in terms of its ability to capture financial return finite-sample properties. Thus, the finite-sample properties and their consequences especially for relatively small samples that match the investment horizon of risk managers need exploration.

The practical implementation of the theory simplifies into developing volatility estimators using aggregated absolute returns, $\sum |r_m|$ and its' variants for any day t with m intraday intervals:

$$|r_t| = \sum_{j=1}^{m} |r_{m, t+j/m}| \qquad (2)$$

For n = 2, this represents the quadratic variation result where squared returns are equated to integrated volatility.

The number of intervals chosen is asset dependent impacted on by such factors as levels of trading activity and of inherent volatility. However, there is a trade-off as m increases the precision of realized power variation increases but microstructure effects such as bid-ask bounce increasing at finer intervals can impair the modelling process. This study follows the standard interval choice of 5-minute intervals throughout the trading day.

As well as directly comparing different volatility series using absolute and squared reaslisations the study examines the ability of the respective measures to filter out the time-varying dynamics associated with asset prices. Daily Returns, $r_t$, obtained by aggregating the high frequency intraday returns, $r_{m,t}$, are rescaled by the respective daily volatility series:

$z_t = r_t/\sigma_t$

where the standardised returns series, $z_t$, are obtained from scaling returns, $r_t$, with each of the volatility proxies, $\sigma_t$.

*Characteristics of volatility series:*

Turning to the application of this method we take high frequency prices for the FTSE100 futures contract traded on LIFFE, for a relatively short time frame between January 1, 1999 through June 30, 2000 using the most actively traded delivery month data from a volume crossover procedure. For each 5-minute interval log closing prices are first differenced to obtain each period's return. The full trading day is between 08.35 and 17.35 entailing 107 5-minute intervals. All non-trading periods and holidays are removed resulting in 375 full trading days for analysis.

Daily returns and daily volatility series are generated from aggregating intraday values such as absolute returns and power variations across the trading day. In order to examine the unconditional distributional properties of the daily return and risk measures summary statistics are estimated detailing four distributional moments presented in table 1. A subset of findings for power coefficients between 0.5 and 1.5 are given.[6] Also, some distributional plots for the returns series, and the volatility and standardised returns series with the most attractive distributional characteristics are given in figure 1. The latter series are of particular interest as we are determining the extent to which we can filter out the non-gaussian features of financial returns using the two sets of volatility series.

INSERT TABLE 1 HERE

INSERT FIGURE 1 HERE

The usual finding for the unconditional distribution of financial returns is evident, namely they are leptokurtotic implying too many realisations bunching around the peak and tails of the distribution relative to gaussianity. In particular the distributional plots indicate the fat-tailed characteristic of financial returns with too many large extreme observations relative to a normal distribution.

In table 1 panel B absolute return volatility and squared return volatility are analysed. Absolute return volatility clearly dominates squared return volatility in terms of desirable time series properties. Whilst the coefficients for third and fourth moments of the volatility series with the most attractive distributional characteristics appear similar, squared returns volatility is more prone to outliers as shown by a very long right tail in figure 1. In general absolute return volatility is more closely associated to a normal distribution than squared return volatility at all power transformations.[7]

---

[6] The main distributional inferences are contained within the results in table 1 and figures 1 and 2. Further results for different power coefficients are available on request.

The standardised returns series, rescaling daily returns by the different volatility is presented in panel C. We are interested in determining whether we can obtain gaussian standardised returns and also identify the volatility processes that allow us to achieve this aim. We find a positive outcome to this endeavour if we standardise by absolute volatility only. Thus unconditionally, returns rescaled by absolute return volatility clearly dominate their squared return counterparts in closely approximating gaussian features. A number of the standardised returns series rescaled by absolute returns exhibit no excess skewness and kurtosis and others show a vast improvement in their characteristics. In fact, the fat-tailed property disappears to the extent that platykurtotic features exist. These rescaled series can now give appropriate risk measures that can be extended further, by for example, incorporating the gaussian square root of time multiplier.

In contrast, the standardised returns rescaled by squared return volatility, with the exception of $[z_t] = [r_t]/[r_t^2]^{0.50}$ representing realised standard deviation, still exhibit strong excess skewness and kurtosis. Interestingly this squared return measure, realised standard deviation, is equivalent to absolute return volatility, $|r_t|$, and is equated to unobservable integrated volatility from the theory of realised power variation.

Other squared return volatility series are unable to capture the dynamics of the returns series adequately. For instance, the much-used realised variance is unable to remove the excess kurtosis of the FTSE100 returns series. Thus for relatively small finite samples it is clear that whilst a spectrum of standardised returns using variants of absolute returns allow the risk manager to present conservative and accurate risk measures that adequately model the time-varying dynamics of asset returns this is not the case for their squared return counterparts.

---

7 Logarithmic transformations are also analysed and confirm these findings. Results available on request

The theory of realised power variation asymptotically allows the conditional distribution of volatility to be random but serially dependent and to exhibit the stylized finding for financial data of volatility clustering. Furthermore, the rescaling of the returns series by the different volatility proxies should produce a white noise series devoid of temporal dependence.

To investigate the finite-sample properties of the use of absolute and squared return volatility and their power variations to match the conditional distribution characteristics of financial time series, figure 2 presents time series plots and sample autocorrelation plots for the returns series, volatility and standardised returns series. The overall finite-sample results suggest that whilst the use of squared realisations meets only some of the criteria to adequately model financial returns, aggregated absolute realisations meet all criteria.

INSERT FIGURE 2 HERE

The returns series exhibit time-varying dynamics along with a very large negative return for August 9, 1999 but is essentially white noise with no significant dependence for 20 lags.

As seen in table 1 both volatility series have unconditional distributions that are fat-tailed and in figure 2 both conditional volatility series vary across time and volatility clusters are clearly evident for the absolute returns series. Volatility clustering is less evident in the squared returns volatility series as a large outlier dominates it on August 9 resulting in a single day's volatility that is more than six times the size of the next largest realisation. Furthermore, the memory of the volatility series using absolute realisations indicates strong serial correlation although no such dependence is evident from using squared realisations, as these are also white noise. Absolute return volatility thus matches the stylized features of financial time series.

*Minimum capital requirements:*
Risk managers are interested in the end product of market risk measures such as minimum capital requirements. The methods outlined for obtaining volatility and standardised returns are now used in to calculate these market risk measures. These capital reserves protect investors against losses arising from the volatility of their holdings and thus adequate modeling of volatility is paramount to their accurate measurement.

Rather than using returns series that would entail an underestimation of risk measures assuming normality, the gaussian standardized returns are analysed. This allows for conservative and consistent risk management estimates. These are presented so as to cover price movements at various probability levels. To illustrate, taking a long position and expressing the minimum capital requirement $Lr_{mincap}$ as a percentage of total investment that covers losses $Lr_{loss}$ at a certain probability:

$$P[L_{loss} < Lr_{min\,cap}] = 0.95 \qquad (3)$$

In this case the capital deposit covers 95% of price movements and losses in excess of this would occur with a 5% frequency. A one-day forecast of the capital required as a percentage of total investment uses chosen quantiles of the standardized returns updated with realized volatility measured by

$$\lambda_t = 1 - \exp(|r_t| z_q) \qquad (4)$$

An illustration of minimum capital requirements for long and short trading positions at common confidence levels is in table 2. For instance, to cover 95% of all price fluctuations in the FTSE100 contract requires a capital deposit of 2.81% of the total investment for a long position. Thus this capital outlay would be insufficient for 5% of the outcomes facing the investor and risk management strategies would be implemented with these capital costs in mind.

INSERT TABLE 2

In conclusion, this paper advocates alternative measures of volatility using aggregated absolute returns and their variations. The measures are underpinned by the theory of realised power variation that asymptotically has absolute variation converging in probability to the unobservable integrated volatility. The practical use of these measures is illustrated in the context of minimum capital requirement estimates, a key market risk measure.

The paper shows that the finite-sample properties of absolute return volatility generally dominate squared return volatility. In particular, rescaling by absolute return volatility results in gaussian standardised returns for a spectrum of power variations. Also, volatility clustering and strong serial correlation are evident for absolute return volatility series matching the properties of financial time series. Moreover, absolute returns are more robust in the presence of extreme returns that result in fat-tails. The key to imposing appropriate risk management measures requires accurate modelling of volatility for different assets. These accurate absolute return volatility measures are used to give conservative daily minimum capital requirements for the FTSE100 futures contract over a small trading period.

References:

Andersen, T. G., T. Bollerslev, and F. X. Diebold (2003). Parametric and nonparametric measurement of volatility. In Y. Ait-Sahalia and L. P. Hansen (Eds.), Handbook of Financial Econometrics. Amsterdam: North Holland.

Barndorff-Nielsen, O. E. and N. Shephard (2003). Realised power variation and stochastic volatility. Bernoulli, 9, 243–265.

Barndorff-Nielsen, O. E. Graversen, S. E., and N. Shephard (2003). Power variation & stochastic volatility: a review and some new results, Unpublished paper: Nuffield College, Oxford.

Brooks, C., A. D. Clare and G. Persand, 2002, Estimating market-based minimum capital risk requirements: A multivariate GARCH approach, Manchester School, 705, 666-681.

Cotter, J., (2004). Minimum Capital Requirement Calculations for UK Futures, Journal of Futures Markets, 24, 193-220.

Davidian, M., & R. J. Carroll (1987). Variance Function Estimation. Journal of the American Statistical Association. 82, 1079-1091.

Karatzas, I., & S. E. Shreve, (1991). Brownian Motion and Stochastic Calculus ($2^{nd}$ ed.). Berlin: Springer-Verlag.

Longin, F.M., (1996). The asymptotic distribution of extreme stock market returns, Journal of Business, 63, 383-408.

Longin, F.M., (2000). From Value at Risk to Stress Testing: The Extreme Value Approach Journal of Banking and Finance 24(7), 1097-1130.

Merton R.C., 1980. On Estimating the Expected Return on the Market, Journal of Financial Economics, 8, 323-361.

Mikosch, T. and C. Starica (2000). Limit theory for the sample autocorrelations and extremes of a GARCH(1,1) process. Annals of Statistics, 28, 1427–1451.

Table 1: Summary statistics for daily FTSE100 series

| | Panel A: Raw Returns | | | | |
|---|---|---|---|---|---|
| Mean | -0.08 | | | | |
| Standard Deviation | 1.34 | | | | |
| Skewness | 0.58* | | | | |
| Kurtosis | 2.64* | | | | |
| | Panel B: Volatility | | | | |
| **Power** | **0.50** | **0.75** | **1.00** | **1.25** | **1.50** |
| | Absolute Returns | | | | |
| Mean | 24.97 | 13.29 | 7.43 | 4.37 | 2.71 |
| Standard Deviation | 3.43 | 2.65 | 2.01 | 1.63 | 1.62 |
| Skewness | 0.11 | 0.63* | 1.12* | 2.23* | 6.52* |
| Kurtosis | 3.72* | 2.42* | 3.12* | 10.55* | 74.08* |
| | Squared Returns | | | | |
| Mean | 7.43 | 2.71 | 1.33 | 1.12 | 1.76 |
| Standard Deviation | 2.01 | 1.62 | 3.23 | 8.69 | 24.06 |
| Skewness | 1.12* | 6.52* | 16.75* | 18.85* | 19.23* |
| Kurtosis | 3.12* | 74.08* | 306.01* | 361.21* | 371.40* |
| | Panel C: Standardised Returns | | | | |
| | Absolute Returns | | | | |
| Mean | 0.00 | 0.00 | 0.00 | 0.00 | 0.00 |
| Standard Deviation | 0.05 | 0.10 | 0.17 | 0.29 | 0.49 |
| Skewness | 0.44* | 0.22 | 0.04 | 0.17 | 0.24 |
| Kurtosis | 2.22* | 1.01* | -0.12 | -0.28 | -0.07 |
| | Squared Returns | | | | |
| Mean | 0.00 | 0.00 | 0.02 | 0.07 | 0.21 |
| Standard Deviation | 0.17 | 0.49 | 1.34 | 3.58 | 9.58 |
| Skewness | 0.04 | 0.24 | 0.46* | 0.86* | 1.30* |
| Kurtosis | -0.12 | -0.07 | 1.35* | 4.63* | 9.25* |

Notes: The daily series are outlined in the text. Normal iid skewness and kurtosis values should have means equal to 0, and variances equal to 6/T and 24/T respectively. Standard errors for the skewness and kurtosis parameters are 0.253 and 0.506 respectively. Significant kurtosis and skewness coefficients are given by *.

Table 2: Minimum capital requirement estimates for daily FTSE100 series

| Probability | 95% | 96% | 97% | 98% | 99% |
|---|---|---|---|---|---|
| Long | 2.81 | 2.96 | 3.01 | 3.40 | 3.95 |
| | [2.53 3.09] | [2.67 3.26] | [2.69 3.33] | [3.74 3.05] | [3.55 4.34] |
| Short | 2.87 | 3.06 | 3.42 | 3.66 | 4.09 |
| | [2.59 3.15] | [2.77 3.36] | [3.10 3.74] | [3.31 4.01] | [3.69 4.49] |

Notes: The minimum capital requirements are expressed as a percentage of the total investment. Results are presented individually for the long and short positions using the methodology outlined in the text. Confidence intervals are given in [].

Figure 1: Distributional plots for daily FTSE100 series

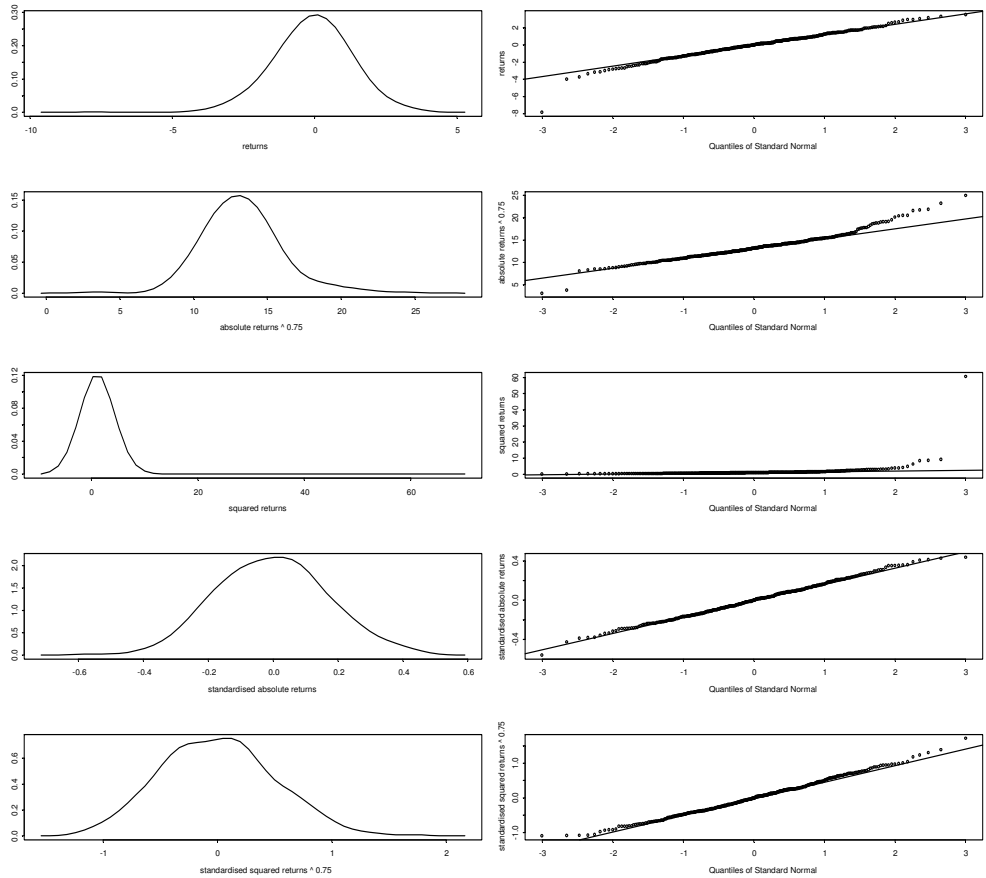

Notes: Density plots followed by q-q plots for the returns, volatility and standardised returns series are presented. The volatility and standardised returns series chosen relying on absolute and squared returns are based on those with the optimal skewness and kurtosis coefficients vis-à-vis normality. Specifically, the volatility series are $|r_t|^{0.75}$ and $[r_t^2]$ and the standardised returns series are $[z_t] = [r_t]/|r_t|$ and $[z_t] = [r_t]/[r_t^2]^{0.75}$.

Figure 2: Time series and Autocorrelation plots for daily FTSE100 series

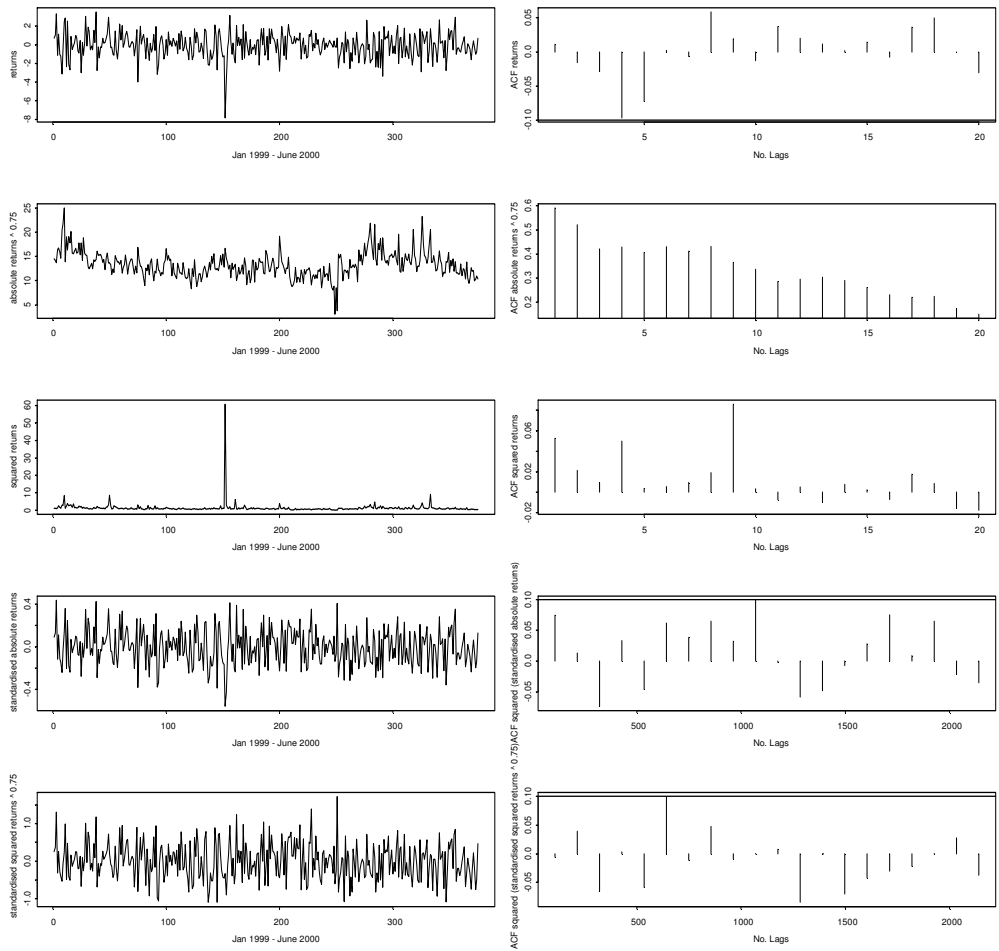

Notes: Time series plots followed by ACF plots for the returns, volatility and standardised returns series are presented. The sample autocorrelations are for a displacement of 20 days from a full sample of 375 days with confidence bands of 0.10. The volatility and standardised returns series chosen relying on absolute and squared returns are based on those with the optimal skewness and kurtosis coefficients vis-à-vis normality. Specifically, the volatility series are $|r_t|^{0.75}$ and $[r_t^2]$ and the standardised returns series are $[z_t] = [r_t]/|r_t|$ and $[z_t] = [r_t]/[r_t^2]^{0.75}$. The ACF plots for the standardised returns series examine squared variations.